\documentclass[notitlepage,twocolumn,prl,tightenlines,nofootinbib,superscriptaddress,showpacs]{revtex4}

\usepackage{amsmath,amssymb,amsfonts,latexsym}
\usepackage{bbm}
\usepackage[mathcal]{euscript}
\usepackage{graphicx}
\usepackage{epsfig}
\usepackage{color}
\usepackage[activate=normal]{pdfcprot}  

\usepackage{psfrag}
\usepackage{txfonts,pxfonts,wasysym,}
\usepackage{pifont}
\usepackage[caption = false]{subfig}

\newcommand{\eccm}{EC2M, UMR Gulliver 7083 CNRS, ESPCI ParisTech,
PSL Research University, 10 rue Vauquelin, 75005 Paris, France}

\newcommand{\be}{\begin{equation}}
\newcommand{\ee}{\end{equation}}
\newcommand{\ben}{\begin{equation*}}
\newcommand{\een}{\end{equation*}}
\newcommand{\ba}{\begin{eqnarray}}
\newcommand{\ea}{\end{eqnarray}}

\newcommand{\bfr}{\mathbf{r}}

\begin{document}

\title{Crystallization of self-propelled hard-discs : a new scenario}

\author{G. Briand}
\affiliation{\eccm}
\author{O. Dauchot}
\affiliation{\eccm}

\date{\today}

\begin{abstract}
We experimentally study the crystallization of a monolayer of vibrated discs with a built-in polar asymmetry, a model system of active liquids, and contrast it with that of vibrated isotropic discs. Increasing the packing fraction $\phi$, the quasi-continuous crystallization reported for isotropic discs is replaced by a transition, or a crossover towards a "self-melting" crystal.  Increasing the packing fraction from the liquid phase, clusters of dense hexagonally-ordered packed discs spontaneously form, melt, split and merge leading to a highly intermittent and heterogeneous dynamics. The resulting steady state cluster size distribution decreases monotonically. For packing fraction larger than $\phi^*$, a few large clusters span the system size and the cluster size distribution becomes non monotonic, the transition being signed by a power-law.
The system is however never dynamically arrested. The clusters permanently melt from place to place forming droplets of active liquid which rapidly propagate across the system. This state of affair remains up to the highest possible packing fraction questioning the stability of the crystal for active discs, unless at ordered close packing.
\end{abstract}

\maketitle
Assemblies of self propelled particles are prone to a number of novel collective behaviors, which are specific to these intrinsically out-of-equilibrium systems~\cite{Marchetti:2013bp,Zottl:2016vi}. As such they open new paths for designing smart materials but also challenge our fundamental understanding of out of equilibrium matter.  

On one hand, the crystallization~\cite{olafsen2005two,reis2006crystallization} and the glass transition~\cite{AbatePRL08,Candelier:2010vo} of mechanically agitated grains, beads or discs, conserve the essential properties of their equilibrium counterparts. Even when, the collisions being significantly inelastic, the 2d crystallization turns into a first order transition with phase coexistence~\cite{Komatsu:2015bh}, it remains an equilibrium concept. A similar result is obtained in a model of repulsive active Brownian particles (ABP)~\cite{Bialke:2012cw}.
But, on the other hand, there are indications that the dense phases of active matter cannot so easily be mapped onto equilibrium situations. The transition shift to higher densities~\cite{Henkes:2011ed,Bialke:2012cw,Fily:2013uu,Ni:1556239,Levis:2014ux,Berthier:2013wg,Szamel:2016er} cannot be explained by a simple scaling argument, using effective temperature. Active glasses exhibit very peculiar structural heterogeneities~\cite{Ni:1556239}. Their dynamics is slower at short times, but faster at large times suggesting that the system is effectively "cooler" than its equilibrium counterpart but also that it access relaxation pathways, which are closed at equilibrium~\cite{Levis:2014ux,Berthier:2013wg}. These observations point at a strong decoupling between structure and dynamics, as also underlined in~\cite{Szamel:2016er}. 
Wether a simple, yet real, system of active particles crystallizes following an equilibrium scenario, remains an open question of both fundamental and practical interest. 


\begin{figure}[t!]
\vspace{-5mm}
\subfloat[]{
\includegraphics[width=0.49\columnwidth,trim = 35mm 10mm 35mm 5mm, clip]{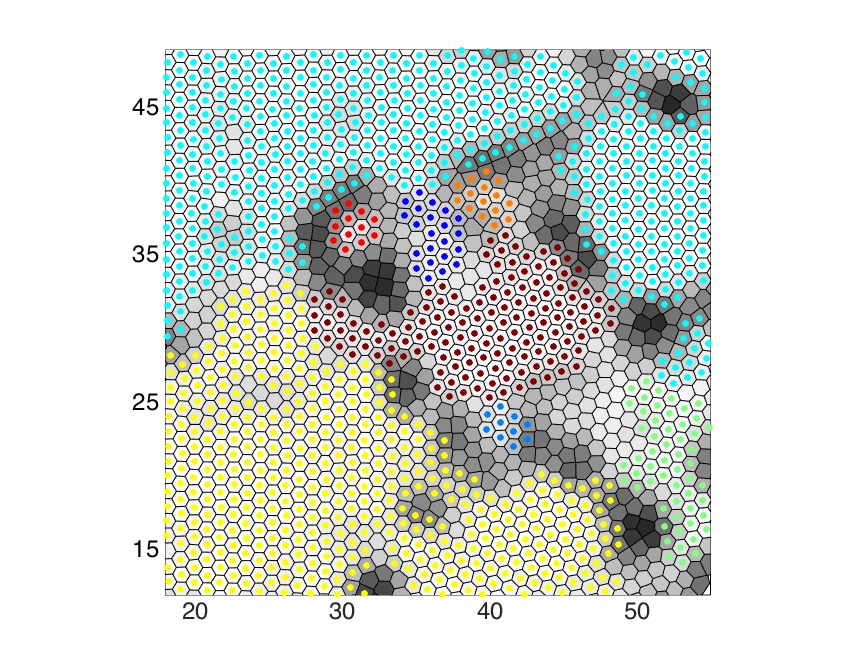}
}
\subfloat[]{
\includegraphics[width=0.49\columnwidth,trim = 35mm 10mm 35mm 5mm, clip]{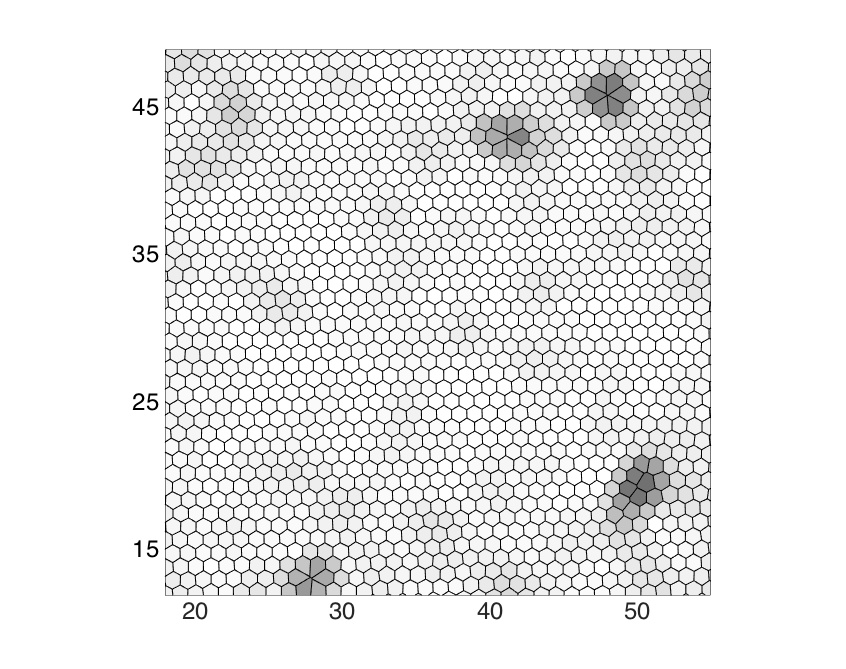}}
\vspace{-2mm}
\caption{Crystallization of a vibrated monolayer of polar \textbf{(a)} and isotropic \textbf{(b)} discs. Packing fraction $\phi = 0.84$. The gray colormap indicates the local orientational order parameter $\psi_6^p$ (see text for details). For the polar discs, distinct crystal clusters (in color) are present, while, apart from local defects, an homogeneous ordered phase is observed in the isotropic case.  (See also movies in Supp. Mat. for the dynamics)}
\label{fig:iso-vs-spp}
\vspace{-6mm}
\end{figure}

In this letter we take advantage of a 2D experimental system of self propelled polar discs~\cite{Deseigne:2010gc,Deseigne:2012kn}, for which high packing fractions $\phi$ can be reached, to perform the first experimental study of crystallization in a system of self propelled discs. We identify a radically new scenario, which share no resemblance with the quasi-continuous, equilibrium-like, crystallization observed for isotropic discs nor with a first order like equilibrium phase coexistence.
Increasing the packing fraction from the liquid phase, clusters of dense hexagonally-ordered packed discs spontaneously form, melt, split and merge leading to a highly intermittent and heterogeneous dynamics. 
For $\phi>\phi^*$, a few large clusters span the system size. The system is however never dynamically arrested. Local excitations form and propagate across the system permanently melting the putative crystalline phase. 

The experimental system made of vibrated discs with a built-in polar asymmetry, which enables them to move coherently has been described in details previously~\cite{Deseigne:2012kn}. The polar particles are micro-machined copper-beryllium discs (diameter $d = 4$ mm) with an off-center tip and a glued rubber skate  located at diametrically opposite positions (total height $h = 2$ mm).  These two "legs", with different mechanical response, endow the particles with a polar axis. Under proper vibration, the discs perform a persistent random walk, the persistence length of which is set by the vibration parameters. We also use plain rotationally-invariant discs (same metal, diameter, and height), hereafter called the ``isotropic'' discs. 
Here we use a sinusoidal vibration of frequency $f=95$ Hz and relative acceleration to gravity $\Gamma = 2\pi a f^2/g = 2.4$. The motion of the particles is tracked using a standard CCD camera at a frame rate of $25$ Hz. In the following, the unit of time is set to be the inverse frame rate and the unit length is the particle diameter. Within these units, the resolution on the position $\vec{r}$ of the particles is better than $0.05$, that on the orientation $\vec{n}$ is of the order of $0.05$ rad. In the present case, the vibration conditions are such that the persistence length of an isolated polar particles $\xi\simeq 5$, is two to three times smaller than in~\cite{Deseigne:2010gc}; no collective motion sets in and the system is closer to existing models, for which the dynamical rules guarantee self-propulsion without alignment~\cite{Lam:2015bp}.
In the following particle trajectories are tracked within a region of interest (ROI) of diameter $20$, where the long-time averaged density field is homogeneous. The average packing fractions $\phi$ measured inside the ROI ranges from $0.42$ to $0.84$.

Since the discovery of the liquid-solid transition for hard disks~\cite{ALDER:1962iz}, the nature of this transition has been a matter of intense debate, until recently~\cite{Bernard:2011bc}, when it was shown that the transition occurs with two steps as in the KTHNY scenario~\cite{Kosterlitz:1973xp,Halperin:1978hw,Strandburg:1988zz}, but with the first transition between the liquid phase and the hexatic phase -- with orientational but no translational order -- being weakly discontinuous. Here also, the transition observed for the isotropic particles follows this quasi-continous scenario, with an homogeneous increase of both $\rho(\bfr)$ and $\psi_6({\bfr})$, when the packing fraction $\phi >\phi^{\dagger}\simeq 0.71$. We leave aside the detailed investigation of this now well characterized transition to concentrate on the case of the polar particles, our model system for self-propelled particles.

\begin{figure}[t] 
\center
\vspace{0cm}
\includegraphics[width=0.48\columnwidth]{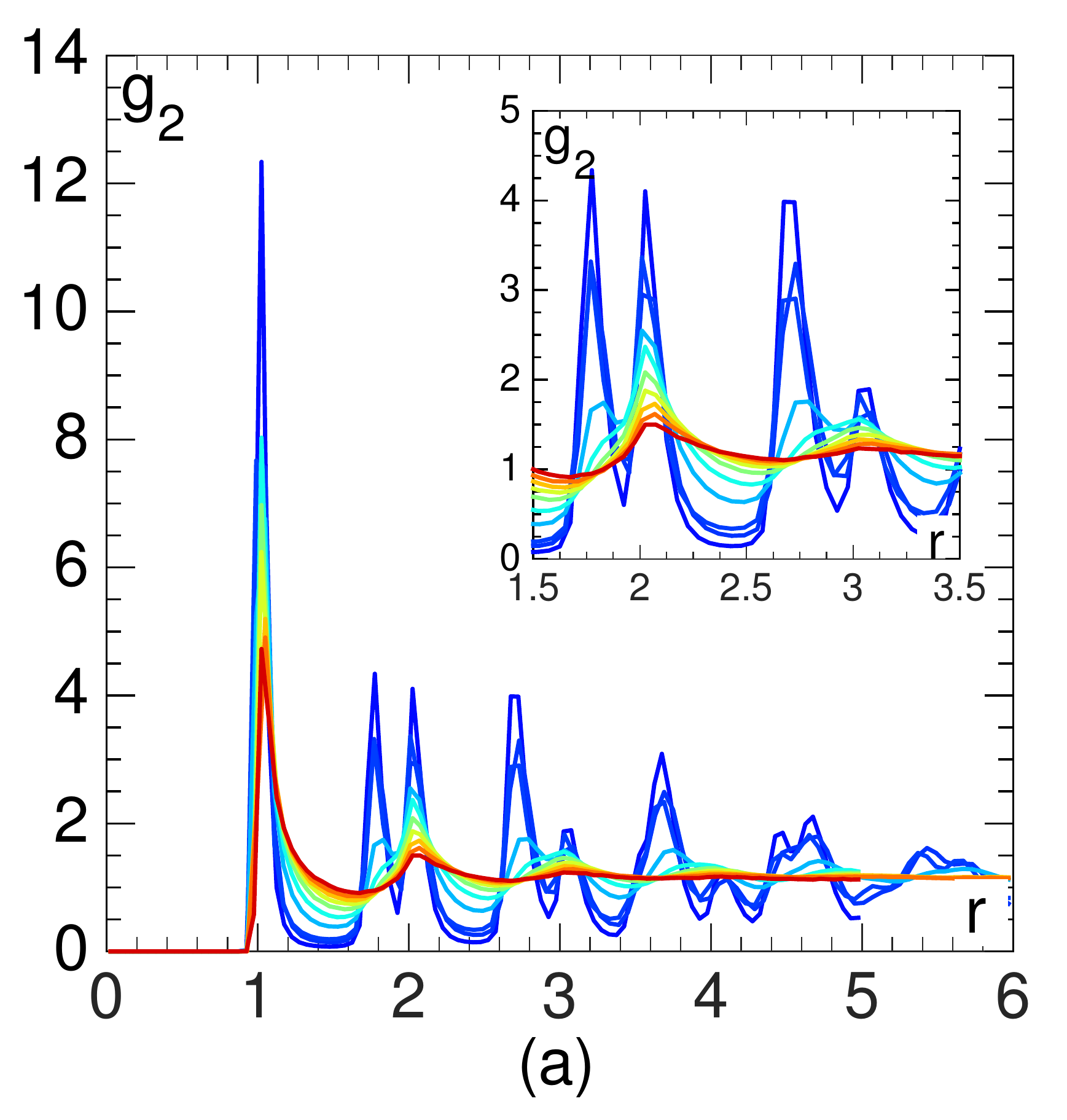}
\includegraphics[width=0.48\columnwidth]{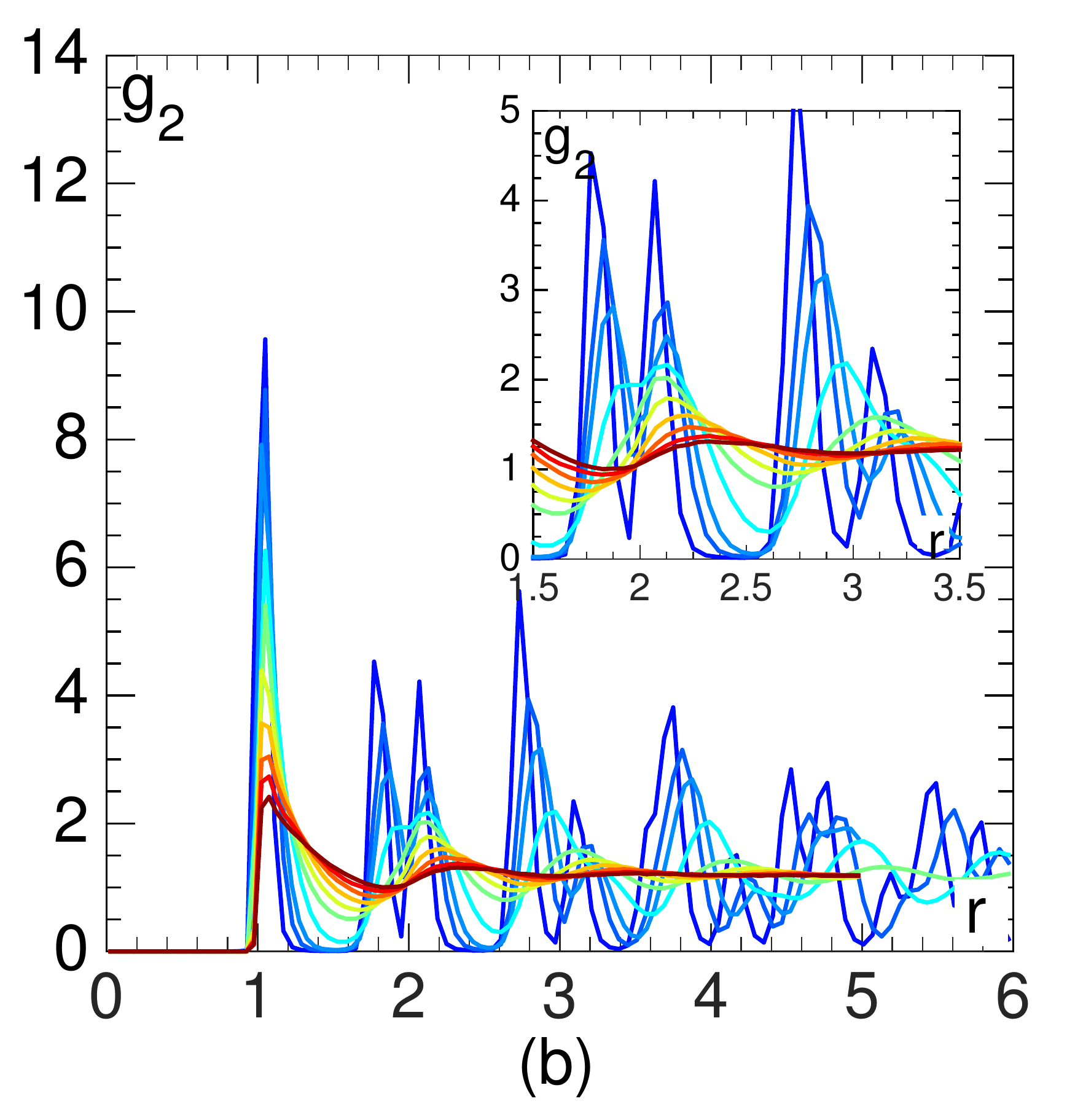} \\
\vspace{-2mm}
\includegraphics[width=0.48\columnwidth]{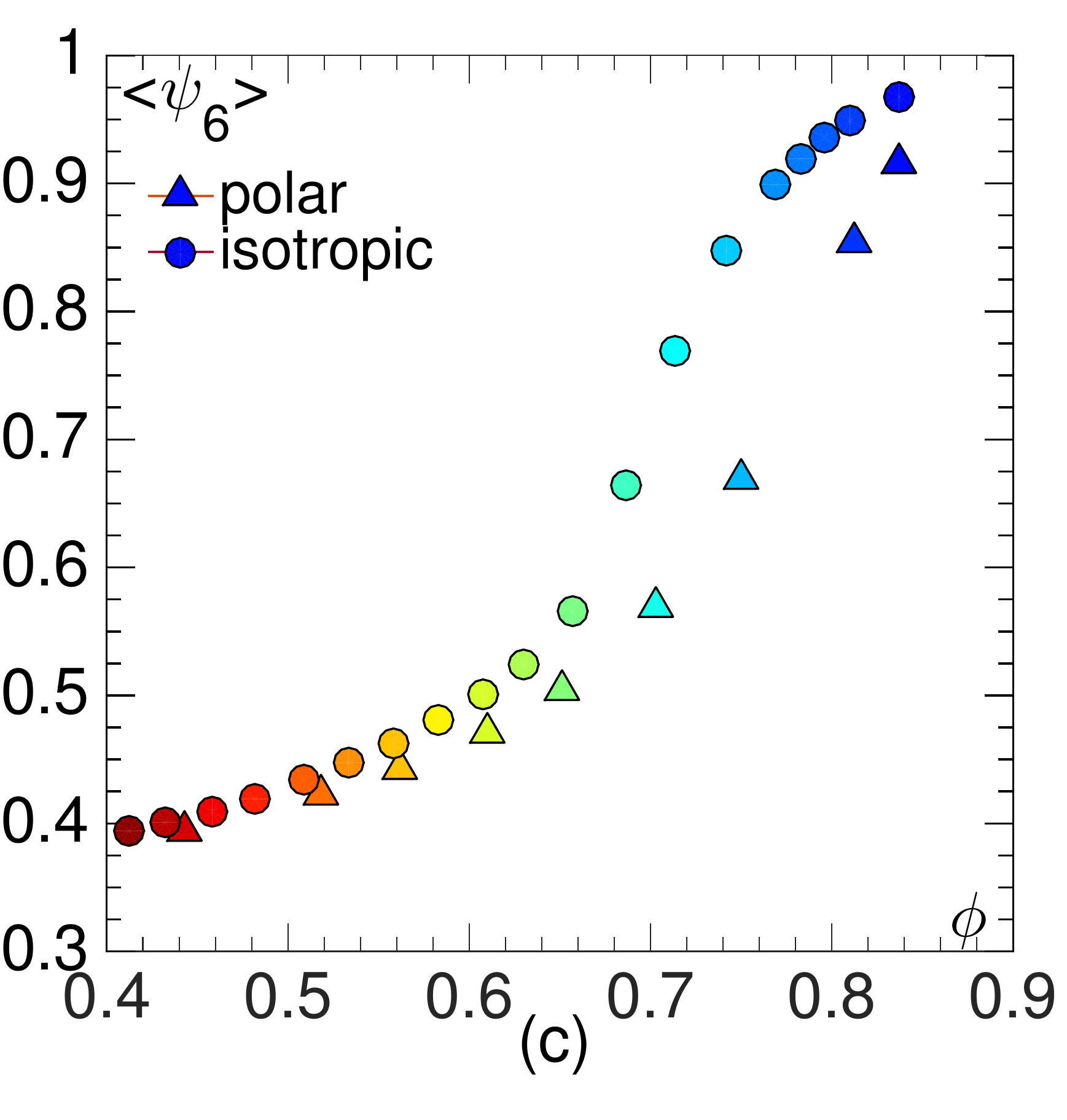}
\includegraphics[width=0.48\columnwidth]{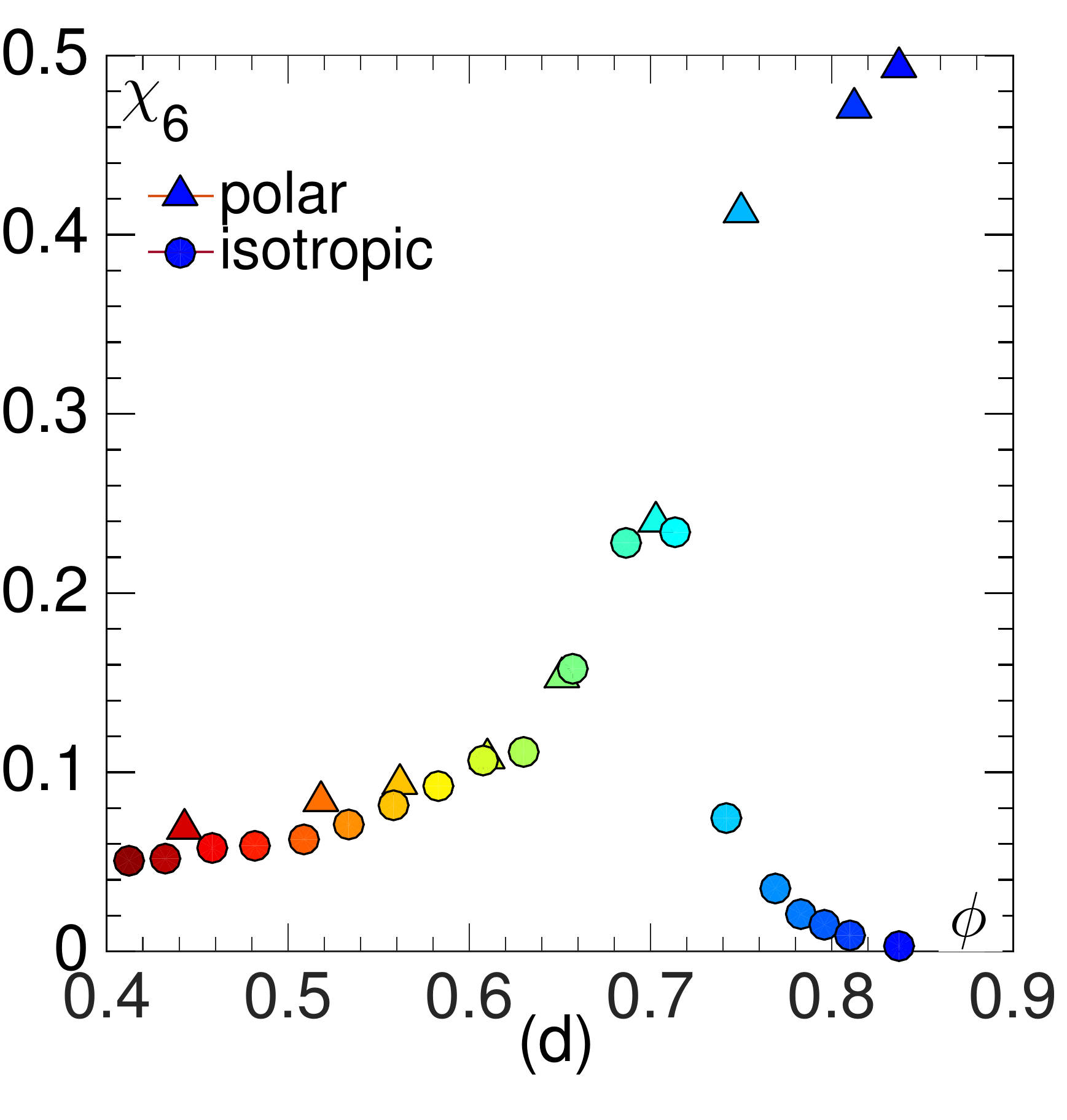}\\
\vspace{-2mm}
\includegraphics[width=0.48\columnwidth]{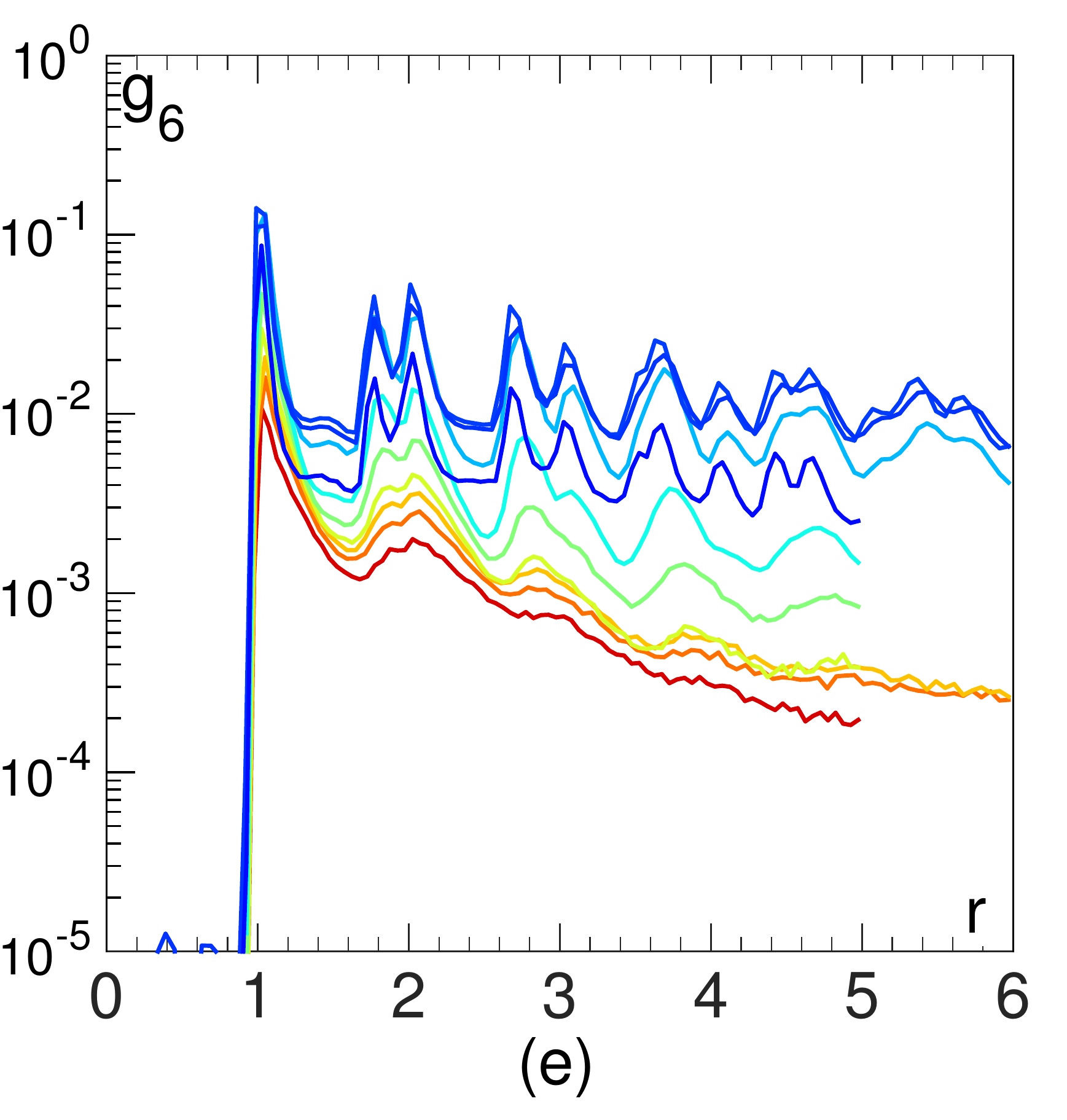}
\includegraphics[width=0.48\columnwidth]{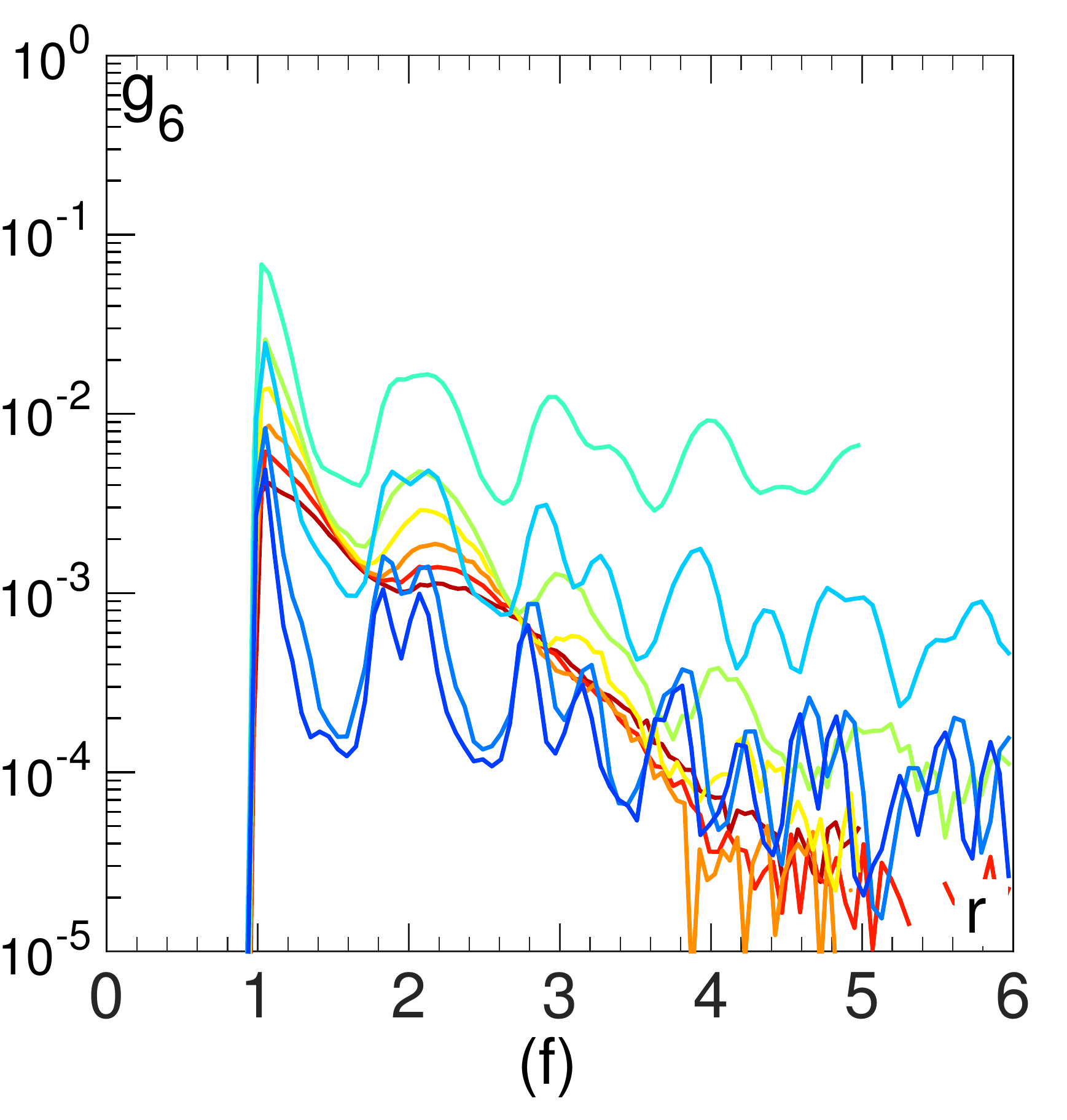}\\
\vspace{-4mm}
\caption{{\bf Structural properties} for $\phi\in [0.42-0.84]$ color coded from red to blue. {\bf (Top):} Pair correlation function for the polar \textbf{(a)} and isotropic \textbf{(b)} discs; (inset: zoom on second and third peaks). \textbf{(Middle):} Dependance on $\phi$ of the mean orientational order parameter $\left<\psi_6\right>$ \textbf{(c)} and its fluctuations \textbf{(d)}; \textbf{(Bottom):} Spatial correlation of $\psi_6$ for the polar \textbf{(e)} and isotropic \textbf{(f)} discs.}
\label{fig:struct}
\vspace{-5mm}
\end{figure}

The structure of the bi-dimensional packing is characterized using standard equilibrium tools. 
Starting from the particle positions at all time $\bfr_p(t)$, we compute the density field $\rho(\bfr)$ and its fluctuations as characterized by the pair correlation function $g_2(r)$:
\be
g_2(r) = \left<\frac{\sum_{p\neq q} \delta\left(r - |\bfr_{q}-\bfr_{p}|  \right)}{2 \pi N r}\right>,
\ee
where $N$ is the number of particles within the ROI at time $t$, and $<\cdot>$ denotes the time average. We also compute the instantaneous orientational order parameter $\psi_6$ at the particle scale, its fluctuations and their correlations $g_6(r)$:
\ba
\psi_6^p & = & \left[\frac{1}{n_p}\sum_{<pq>}\exp(6i\theta_{pq})\right], \\
g_6(r) & = & \left<\frac{\sum_{p\neq q} \psi_6^p \psi_6^{q} \delta\left(r - |\bfr_{q}-\bfr_{p}| \right)} {2 \pi N(N-1) r}\right>
\ea 
where $\sum_{<pp'>}$ denotes the sum over the $n_p$ neighbors of particle $p$ identified from a Vorono\"{i} tessellation, and $[\cdot]$ a coarse-graining of the field on the first neighbors shell.

Figure~\ref{fig:struct} synthesizes the structural properties of the polar discs system and how they compare with the case of the isotropic discs. The pair correlation function (fig.~\ref{fig:struct}(a)) clearly exhibits the signature of an emerging crystal structure for packing fractions similar to that of the polar discs. However a closer examination indicates that the location of the secondary peaks coincide with that of the hexagonal close packing (HCP) as soon as they develop, in sharp contrast with the isotropic case (fig.~\ref{fig:struct}(b)), for which the peaks progressively shift to the right when further compressing the crystal formed at $\phi^{\dagger}$. The structures forming in the system of polar particles are densely packed hexagonally ordered clusters. 
Examining the statistics of $\psi_6 = \frac{1}{N}\sum_p \psi_6^p$ the orientational order parameter further confirms this observation (fig.~\ref{fig:struct}(c-d)). 
In the case of the polar particles, the temporal average $\left<\psi_6 \right>$ and temporal fluctuations, also called the susceptibility $\chi_6 = N \text{var} (\psi_6)$ smoothly increase with the packing fraction. There is no inflection in $\left<\psi6 \right>(\phi)$ and no maximum in $\chi_6(\phi)$, as observed in the case of the isotropic particles. 
This behavior reflects that for the polar discs the probability distribution function (pdf) of $\psi_6$ (not shown here) display a bimodal shape, which is absent in the case of the isotropic discs. These observation all take their roots in the fact that the spatial correlation continuously grow, suggesting the existence of larger and larger domains, in contrast with the case of the isotropic disks for which  the spatial correlations of $\psi_6$ exhibit a non monotonic dependance on $\phi$, with a characteristic length scale that is maximal close to $\phi^{\dagger}$ (fig.~\ref{fig:struct}(e-f)).

The structural analysis reveal that the emergence of crystal order in the polar discs system follows a very different scenario from the one reported at equilibrium or for the isotropic discs. A coexistence picture, suggestive of a first order transition, replaces that of a quasi-continuous transition. Turning to the study of the dynamics, we shall see however that no part of the system ever freeze so that this picture is not correct either.

\begin{figure}[t] 
\center
\vspace{0cm}
\includegraphics[width=0.48\columnwidth]{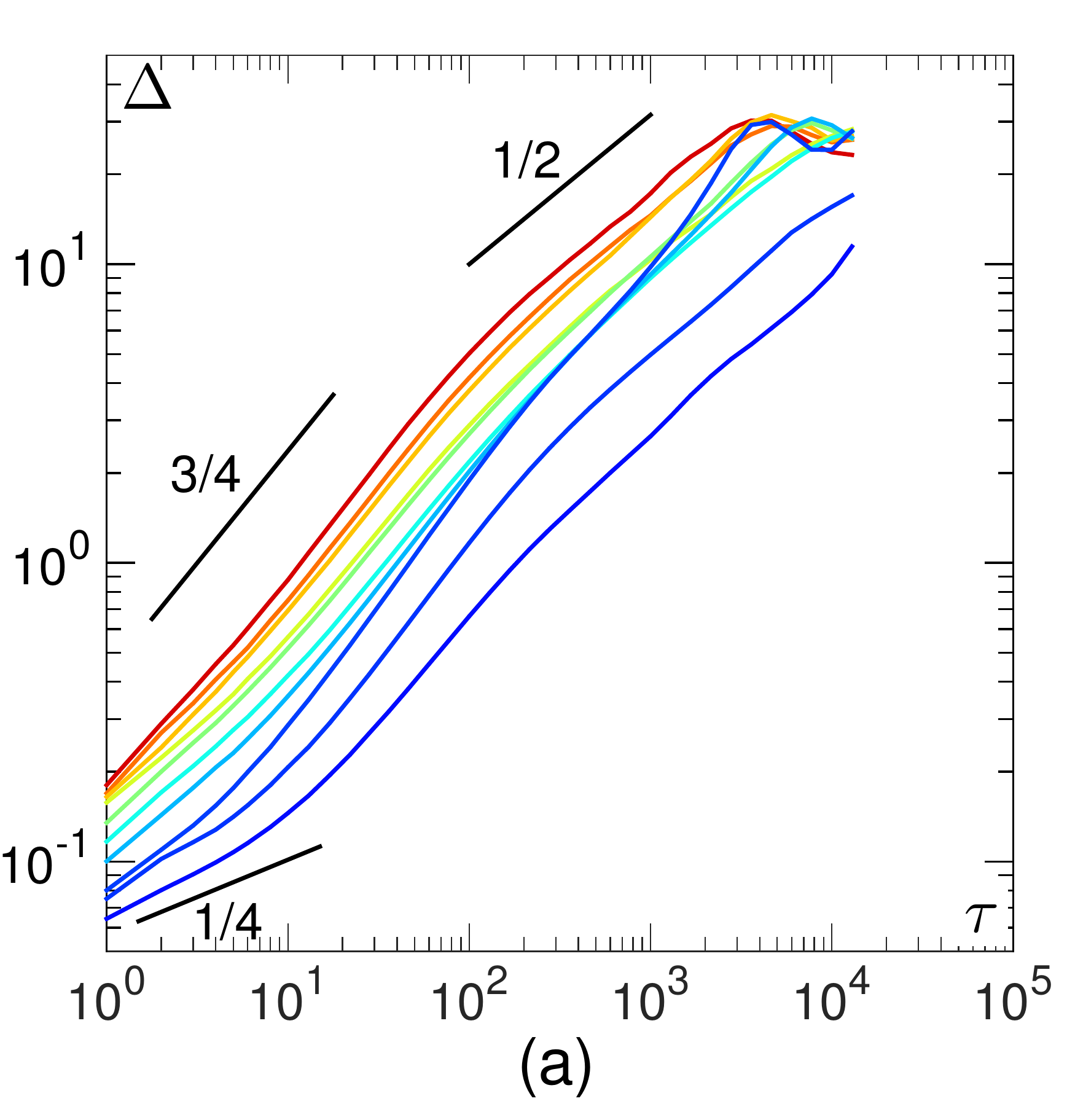}
\includegraphics[width=0.48\columnwidth]{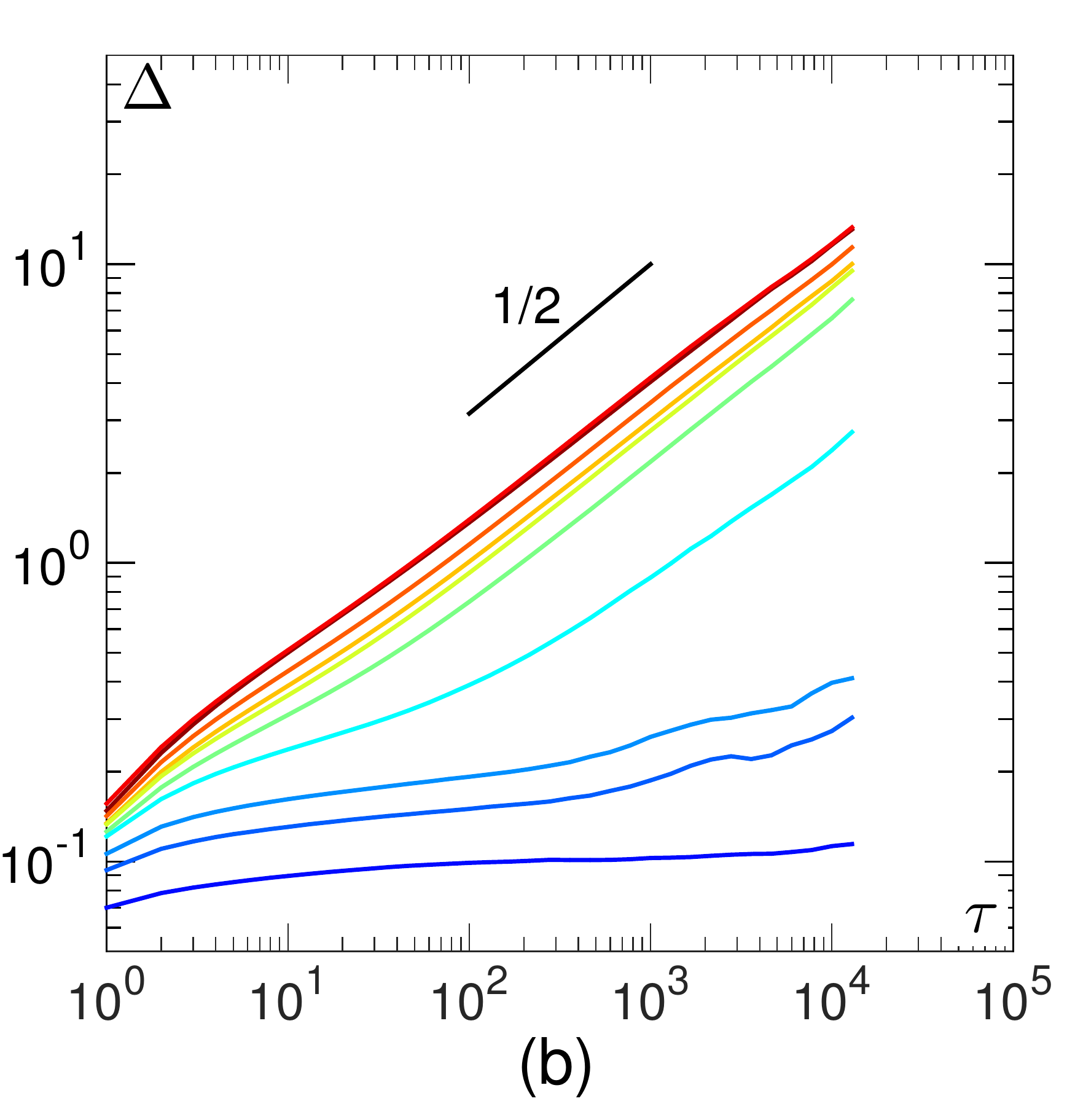} \\
\vspace{-2mm}
\includegraphics[width=0.48\columnwidth]{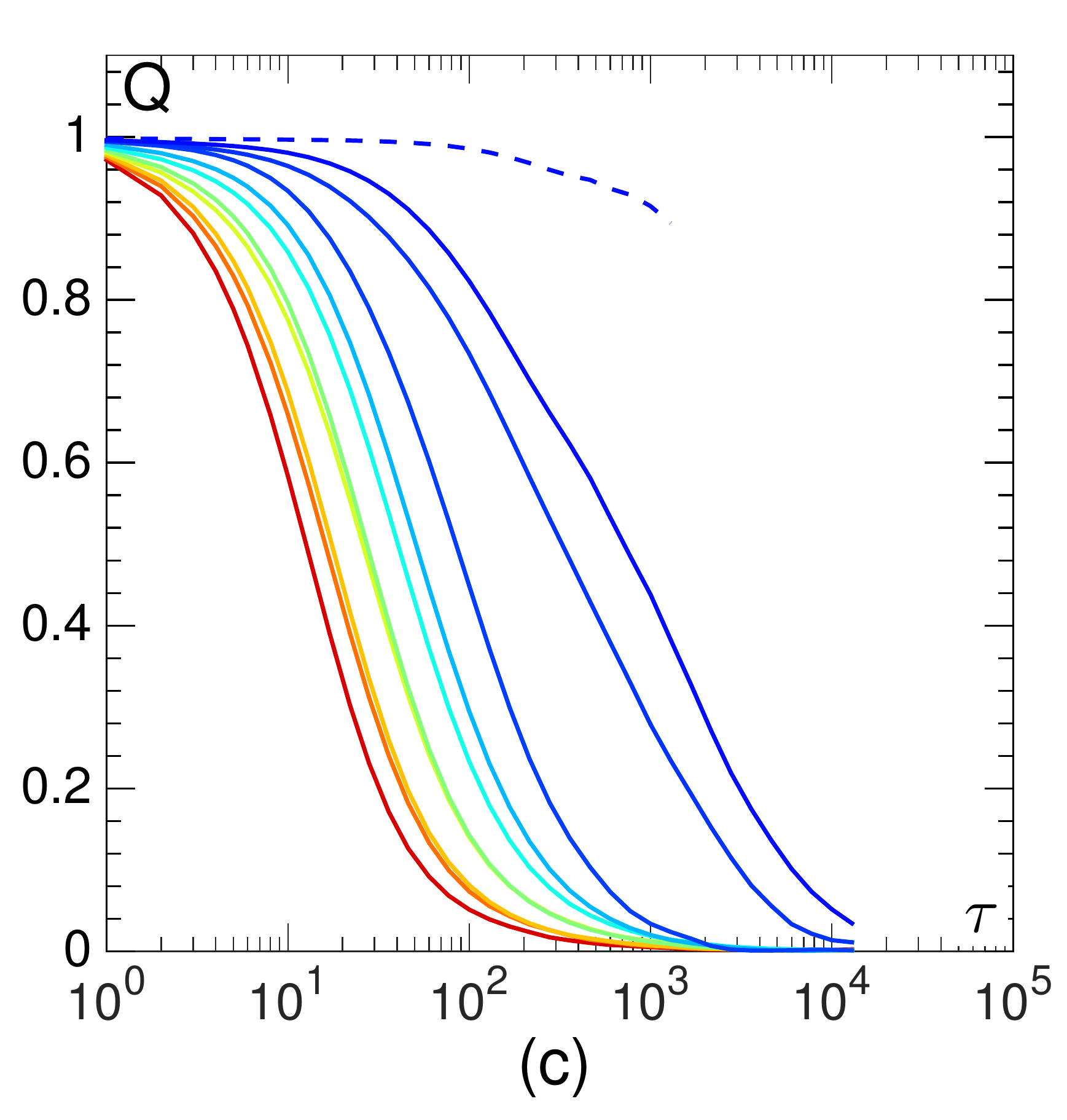}
\includegraphics[width=0.48\columnwidth]{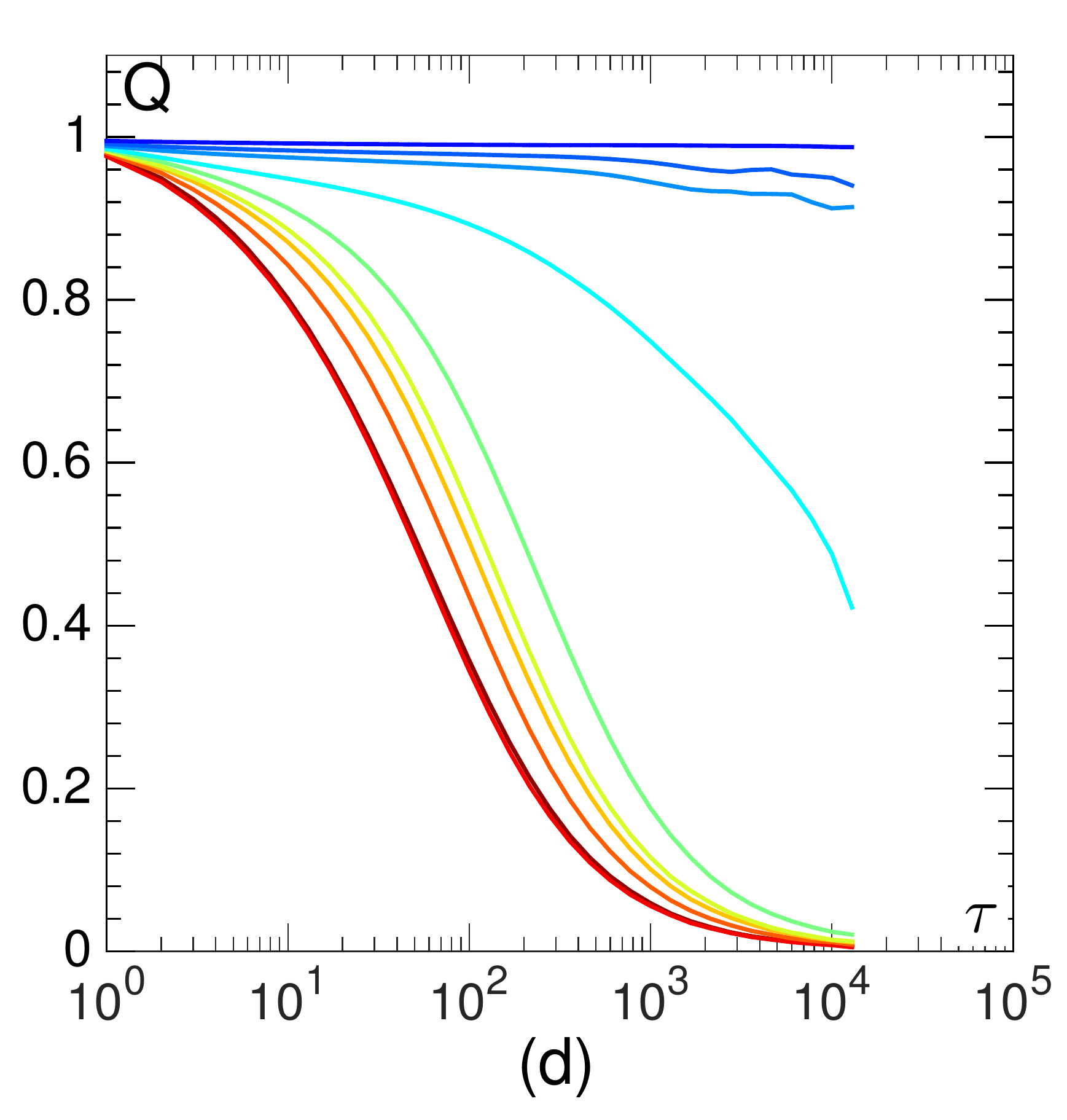}\\
\vspace{-2mm}
\includegraphics[width=0.48\columnwidth]{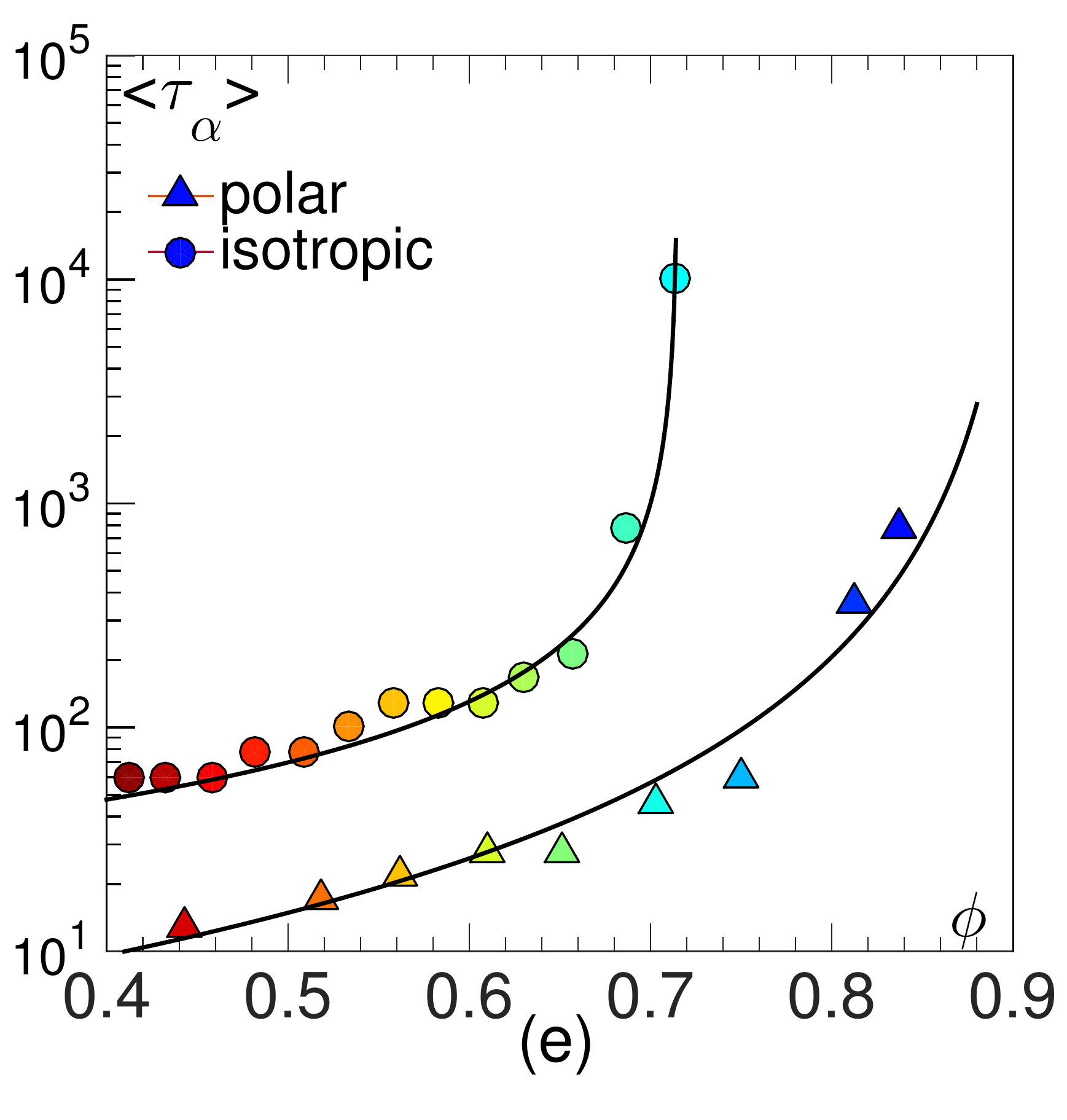}
\includegraphics[width=0.48\columnwidth]{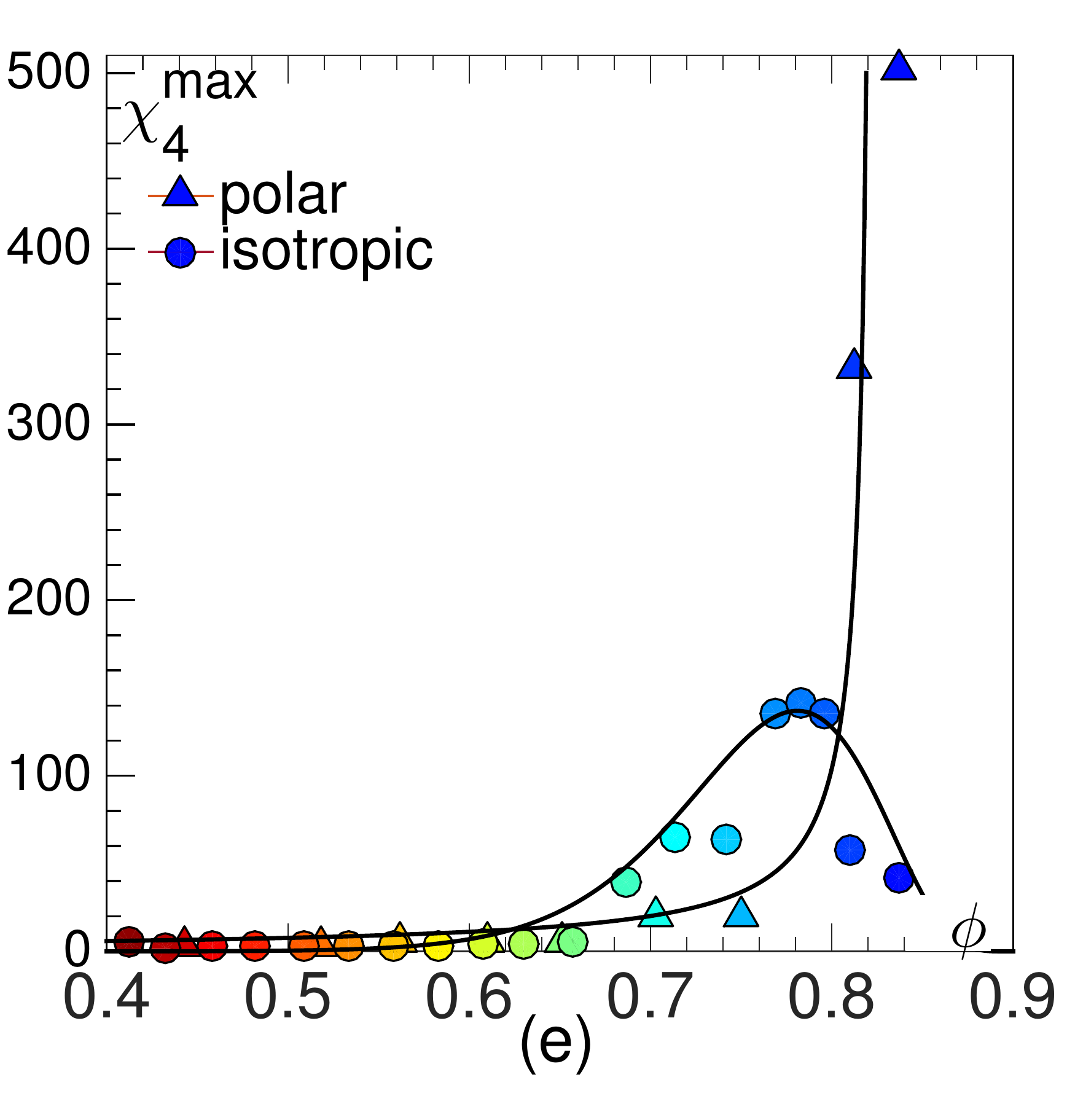}\\
\vspace{-4mm}
\caption{{\bf Dynamical properties.} Mean square displacement {\bf (Top)} and self-part of the dynamical overlap function \textbf{(Middle)} for different $\phi$ for the polar \textbf{(a),(c)} and isotropic \textbf{(b),(d)} discs. The dotted line in \textbf{(c)} shows the relaxation of particles included in a crystalline cluster (see text for details). {\bf (Bottom):} Relaxation time $\tau_\alpha$ \textbf{(e)} and maximal dynamical susceptibility $\chi_4^{\rm max}$ \textbf{(f)} as a function of $\phi$. Same color code as in fig.~(\ref{fig:struct}). }.
\label{fig:dyna}
\vspace{-0.5cm}
\end{figure}

The mean square displacement (MSD) $\Delta^2(\tau) = \left<\frac{1}{N}\sum_p \left(\bfr_p(t+\tau)-\bfr_p(t)\right)^2 \right>$ of the polar particles is super-diffusive until $\tau = 100$, where normal diffusion sets in, \emph{for all packing fractions} [fig.~\ref{fig:dyna}(a)]. This is in sharp contrast with the case of the isotropic discs [fig.~\ref{fig:dyna}(b)], for which a clear plateau develops above $\phi^{\dagger}$, associated with the trapping of the particles in the crystal structure. As a matter of fact, the short time dynamics of the polar particles, does present a small sign of trapping at the largest $\phi$, but this is rapidly wiped out by the longer term super diffusion. 
The decrease in magnitude of the MSD with increasing $\phi$, could suggest that a larger and larger fraction of the particles are trapped, while the remaining ones behave as an active liquid. This is however not the correct picture as demonstrated by the large-time behavior of the self part of the dynamical overlap function $Q(a,\tau)$ and of the dynamical susceptibility $\chi_4(a,\tau)$~\cite{vanSaarloos:2011wv}:
\ba
Q(a,\tau) &=& \left<\frac{1}{N}\sum_p \exp -\frac{\left(\bfr_p(t+\tau)-\bfr_p(t)\right)^2}{a^2} \right>\\
\chi_4(a,\tau) &=& N \text{var} \left(\frac{1}{N}\sum_p \exp -\frac{\left(\bfr_p(t+\tau)-\bfr_p(t)\right)^2}{a^2} \right)
\ea 
which we evaluate for $a=1$. $Q(\tau)$, instead of developing a finite value plateau, pointing at a fraction of dynamically arrested particles, always rapidly decreases to zero : all particles move more than one diameter on timescales of the order of $5000$ [fig.~\ref{fig:dyna}(c)]; no part of the system is dynamically arrested. By comparison, in the case of the isotropic particles, $Q(\tau)$ clearly converges towards a plateau close to one [fig.~\ref{fig:dyna}(d)] when $\phi>\phi^{\dagger}$. Accordingly, while the relaxation time  $\tau_\alpha$, defined by $Q(\tau_\alpha)=0.5$, diverges sharply at $\phi^{\dagger}$, pointing at the crystallization transition for the isotropic particles, it middy increases for the polar ones [fig.~\ref{fig:dyna}(e)]. 
%
\begin{figure}[t] 
\center
\vspace{0cm}
\includegraphics[width=0.48\columnwidth]{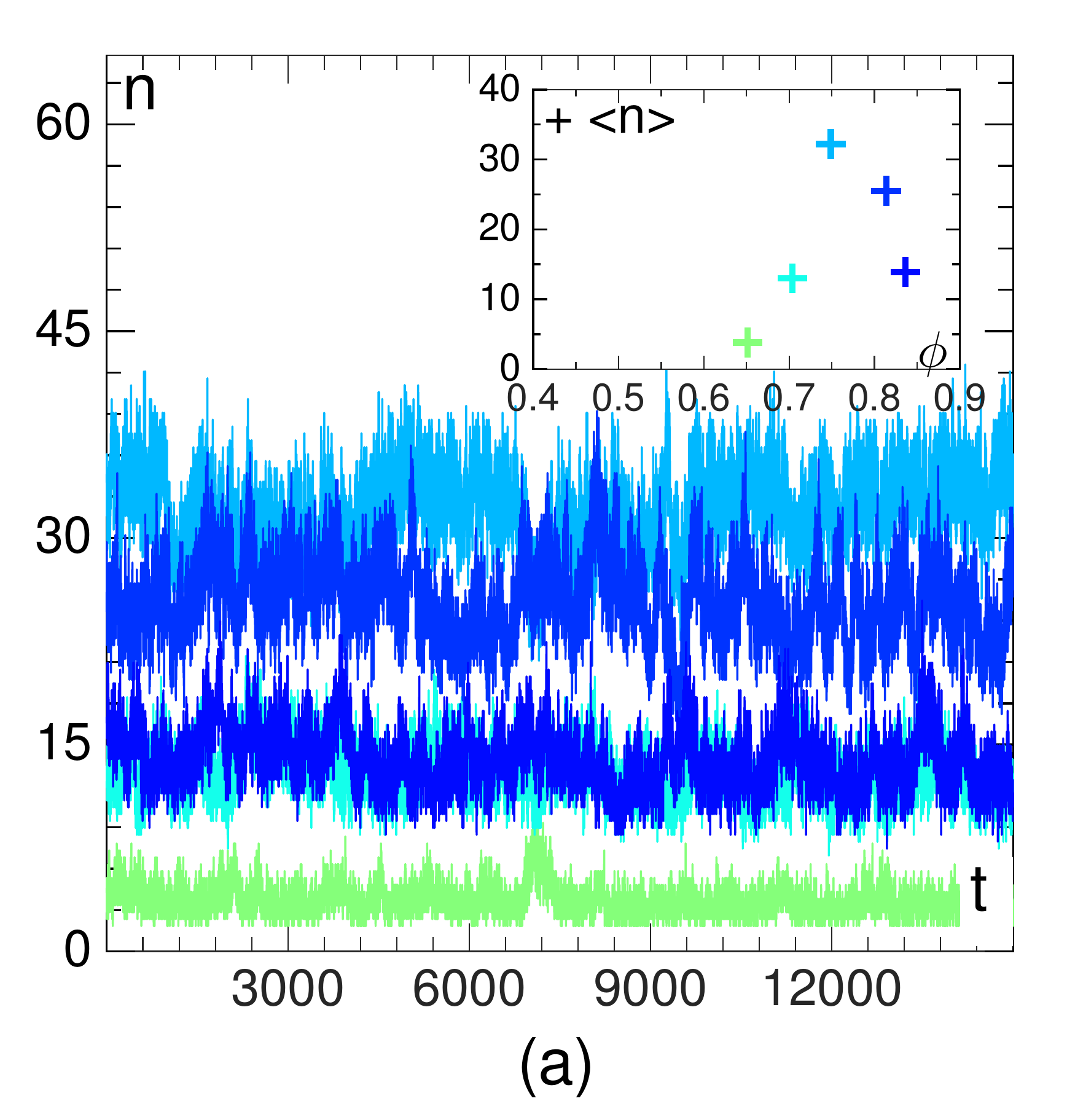}
\includegraphics[width=0.48\columnwidth]{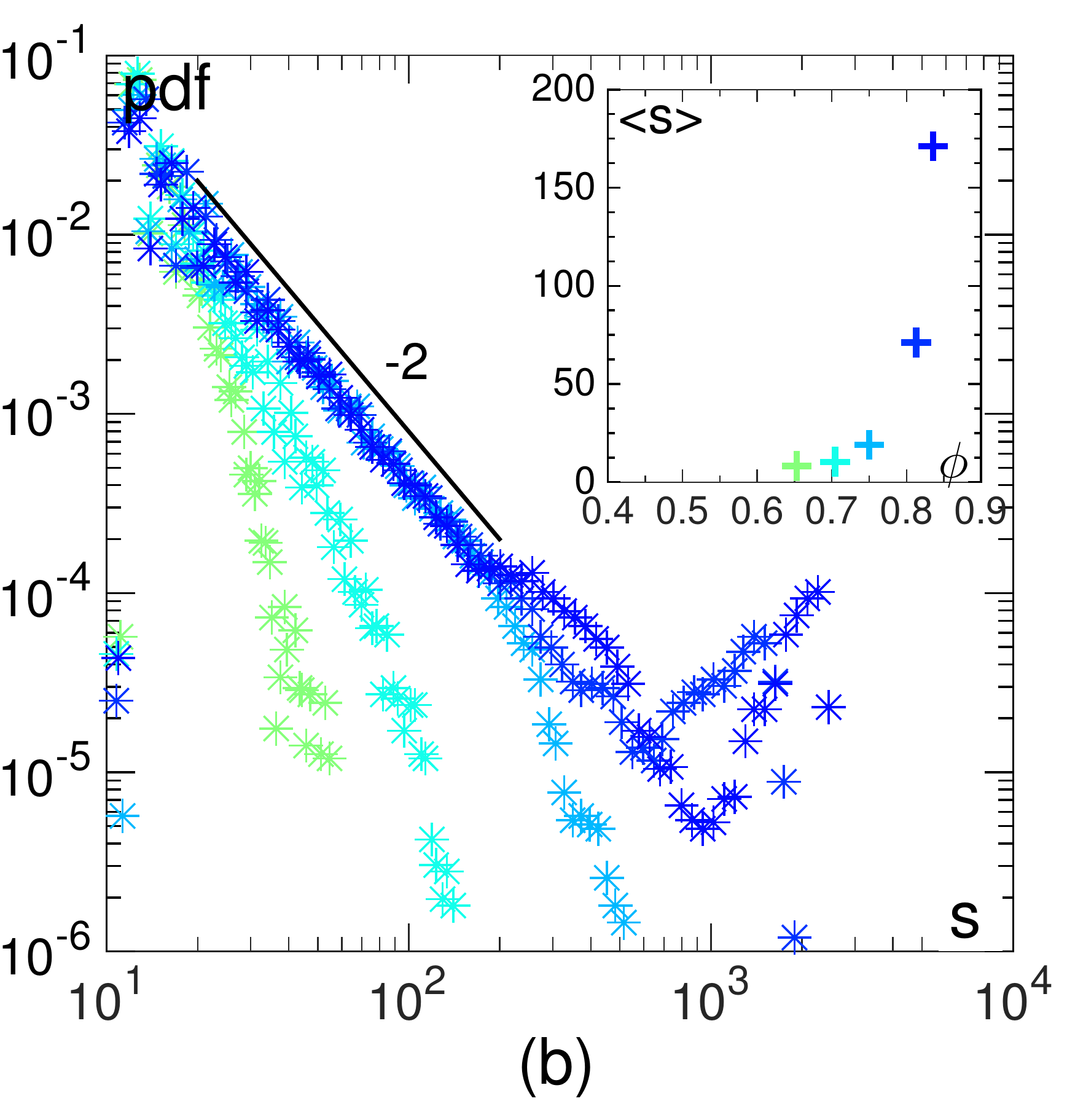} \\
\vspace{-3mm}
\caption{{\bf HCP clusters statistics} \textbf{(a):}Steady evolution of the number of clusters (inset: average number of cluster vs. $\phi$), and \textbf{(b):} cluster sizes distributions (inset: average cluster size vs. $\phi$), for the five largest packing fractions where clusters appear.}
\label{fig:clusters}
\vspace{-0.6cm}
\end{figure}
The maximum of the dynamical susceptibility, $\chi_4^{max} = \text{max}(\chi_4(\tau))$, which takes place for $\tau\simeq\tau_\alpha$, and quantifies the heterogeneities of the dynamics, exhibits a mild maximum in the transitional regime for the isotropic particles, while it becomes increasingly large when entering the coexistence regime for the polar particles [fig.~\ref{fig:dyna}(f)]. For the isotropic particles, the dynamical heterogeneities reflects the structural ones : they gently increase in the transitional regime, but disappear once in the homogeneous crystalline phase. The case of the polar particles is more intriguing : not only the dynamical heterogeneities increase continuously with the packing fraction; they also increase much faster than the relaxation time, pointing at a peculiar collective behavior at some intermediate packing fraction $\phi^{*}\simeq 0.82$. 

Further insight into this unexpected feature comes from a closer inspection of the densely ordered clusters unveiled by the structural analysis (see also Movies in Sup Matt.). A cluster is defined as a group of particles sharing six neighbors "in contact" ($0.9d<r_{ij}<1.1d$). By convention, the neighbors are also included in the cluster. The number of clusters (fig.~\ref{fig:clusters}-a), fluctuates around a steady value, with no sign of coarsening, at all packing fraction. The average number of cluster is maximum for $\phi\simeq\phi^*$. 
For $\phi<\phi^*$, clusters split and merge leading to a steady distribution of cluster sizes $p(s)$ essentially decreases exponentially (see fig.~\ref{fig:clusters}-b). As the packing fraction increases towards $\phi^*$, the distribution approaches a power-law $p(s)\sim s^{-\gamma}$, $\gamma = 2$, with a system size cut-off. For 
$\phi>\phi^*$ it is non-monotonic and a peak at large cluster sizes emerge. This behavior is reminiscent of a transition reported in several experiments with bacteria~\cite{Zhang:2010jn,Chen:2012fs,Peruani:2012dy} and simulations~\cite{Yang:2010iq,Peruani:2012dy}. Also, the value of $\gamma=2$, is very close~\footnote{$p(s)$, the cluster size distribution is easily related to $\tilde p(s)$, the probability of a particle to be in a cluster of size $s$: $\tilde p(s) \propto s p(s)$} to the one obtained in simulations of self-propelled rods $\gamma = 1.9$, in experiments on myxobacteria $\gamma = 1.88$ and compatible with that obtained from simplified kinetic models of cluster dynamics~\cite{Peruani:2013ec,Levis:2014in}. We note that the present observation of a phase of dynamical clusters demonstrates that diffusio-phoretic sensing is not necessary for this phase to take place.
Even the largest clusters which form for $\phi>\phi^*$ and span the system size are never dynamically frozen : the locally ordered structure spontaneously melts (see also Movie in Sup Matt Movie) leading to the intermittent formation of active droplets rapidly propagating and relaxing the system. 

The above scenario suggests that no crystal phase stabilizes below HCP. To further confirm this observation, we 
compute the dynamical overlap function $Q(\tau)$ for a set of particles which remain at all times inside the longest lived cluster, at the largest packing fraction $\phi = 0.837$ explored here. The advection of the cluster is removed by computing the particles displacements in the frame of their center of mass. Doing so, we evaluate the relaxation time of the polar particle crystalline state if any. The result is displayed in dashed line on fig.~\ref{fig:dyna}(c): (i) the relaxation is faster than that of the passive crystal at the same packing fraction, indicating internal relaxation processes, much faster than equilibrium defects dynamics; (ii) we could not computed the relaxation on longer time scales, because of the cluster splitting into pieces.  Investigating the local melting processes at play in detail is beyond the scope of the present study. Visual inspection however suggest two complementary mechanisms. On one hand the polar discs tendency to cluster at the highest possible packing fraction frees some volume, where melting can take place. On the other hand active stresses can locally shear and enforce local melting.

{\it --- Discussion ---}
Some of the above conclusions might be related to the confinement and finite size of our experimental system.
As discussed in~\cite{Peruani:2013ec}, depending on the splitting and aggregation processes, the transition can be a guenuine phase transition, which persist in the thermodynamic limit, or just a crossover. In the latter case,  the packing fraction $\phi^*$ would increase with system size, hit the HCP limit and the system would remain in the many-cluster phase at all $\phi$.

We conclude by pointing interesting differences with previous studies. The present scenario is drastically different from that reported for ABP~\cite{Bialke:2012cw} suggesting that ABP and the present vibrated polar discs belong to different classes of self propelled discs.
As compared to the numerical studies~\cite{Ni:1556239,Levis:2014ux}, the dynamical decoupling reported here is extreme in the sense that dynamical arrest only occurs at close packing. Confirming this in the case of a bi-disperse disordered system, would imply the absence of glass transition and finite time relaxation up to jamming.
Finally, classical nucleation theory was recently extended to active systems to describe the aggregation process following the MIPS~\cite{Richard:2016hk,Redner:2016vt}. Our results suggest that alternative, more radically different, approaches might be necessary to deal with the very dense phases of active matter.

{\it --- Acknowledgment ---}
We thank Michael Schindler for helpful suggestions on the clustering algorithm, Ludovic Berthier and Fernando Peruani for valuable comments.
\vspace{-2mm}
\bibliography{/Users/olivierdauchot/Documents/_Science/Biblio/Active.bib,/Users/olivierdauchot/Documents/_Science/Biblio/Crystallisation.bib,/Users/olivierdauchot/Documents/_Science/Biblio/Glasses.bib}


\newpage
{\it --- Supplementary Material ---}
Here we would like to provide further visual evidences of our main observations.
We provide one movie of the cluster dynamics for $\phi\simeq\phi^*$
\begin{itemize}
\vspace{-2mm}
\item Clusters-Polar-phi=0.812.avi
\vspace{-2mm}
\end{itemize}
The black and white field codes $\psi_6$. The color code for the cluster is for identification only.
The movie contains 1000 images shown at 30fps.

We provide 2 movies of the field of $\psi_6$ for comparison of the dynamics in the polar and isotropic dense phases. The packing fraction is the same : $\phi=0.837$ in both cases. The black and white field codes $\psi_6$. The movies contain 1000 images shown at 30fps.

\begin{itemize}
\vspace{-2mm}
\item Psi6-Polar-phi=0.837.avi
\vspace{-2mm}
\item Psi6-Iso-phi=0.837.avi
\vspace{-2mm}
\end{itemize}
The local defects of the crystalline order move around the whole polar system, while they remain localized in the isotropic case. Doing so they allow for the compete relaxation of the system as further illustrated on the figure below, which contrast the structure and the dynamics of the polar and isotropic particles: in the polar case the whole system relaxes rapidly; not only the regions where the defects are at the initial time.

\begin{figure}[h!]
\vspace{0mm}
\includegraphics[width=0.48\columnwidth]{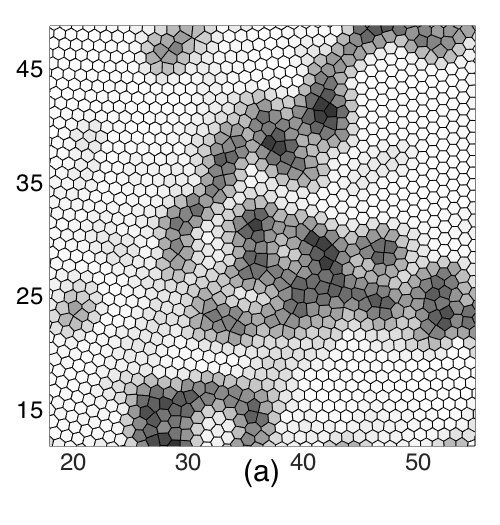}
\includegraphics[width=0.48\columnwidth]{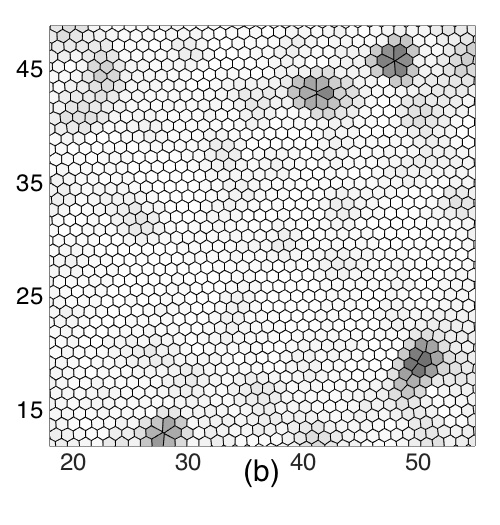}\\
\vspace{-4mm}
\includegraphics[width=0.48\columnwidth]{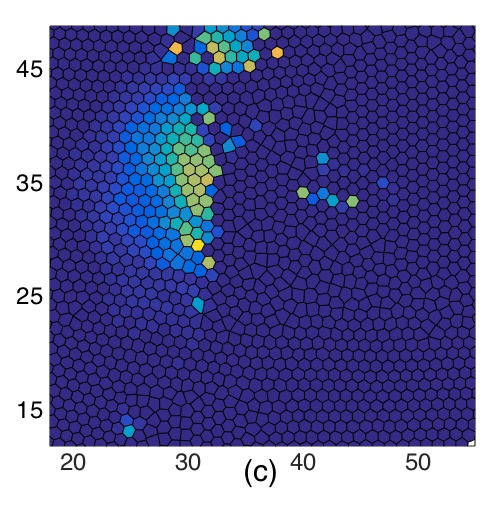}
\includegraphics[width=0.48\columnwidth]{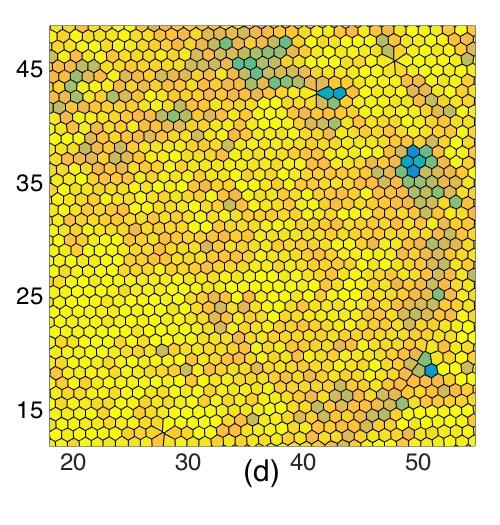}\\
\vspace{-5mm}
\caption{Polar (\textbf{left}) and isotropic (\textbf{right}) discs. $\phi = 0.837$. Top row: $\psi^p_6(t0)$ gray-coded from black(0) to white (1). Bottom row: $Q^p(a=1,t_0,t_0+\tau=500)$, color coded from blue(0) to yellow(1).}
\label{fig:iso-vs-spp}
\vspace{-6mm}
\end{figure}

\end{document}